%% file: main.tex
\documentclass[conference]{IEEEtran}
\usepackage{cite}
\usepackage{amsmath,amssymb,amsfonts}
\usepackage{graphicx}
\usepackage{textcomp}
\usepackage{commath}
\usepackage{wrapfig}
\usepackage{subcaption}
\usepackage{tikz}
\usepackage{hyperref}
\usepackage{fancyhdr}

\newcommand\copyrighttext{%
  \footnotesize
  53rd Asilomar Conference on Signals, Systems, and Computers (ACSSC 2019).
  \textcopyright 2019 IEEE. Personal use of this material is permitted.  Permission from IEEE must be obtained for all other uses, in any current or future media, including reprinting/republishing this material for advertising or promotional purposes, creating new collective works, for resale or redistribution to servers or lists, or reuse of any copyrighted component of this work in other works. DOI: \href{https://doi.org/10.1109/ieeeconf44664.2019.9048906}{10.1109/IEEECONF44664.2019.9048906}}
\newcommand\copyrightnotice{%
\begin{tikzpicture}[remember picture,overlay]
\node[anchor=south,yshift=15pt] at (current page.south) {\fbox{\parbox{\dimexpr\textwidth-\fboxsep-\fboxrule\relax}{\copyrighttext}}};
\end{tikzpicture}%
}

\begin{document}

\title{A Generalized Proportionate-Type Normalized Subband Adaptive Filter}

\author{
\IEEEauthorblockA{\textit{Kuan-Lin Chen, Ching-Hua Lee, Bhaskar D. Rao, and Harinath Garudadri} \\
\vspace{0.25cm}
\IEEEauthorblockA{Department of Electrical and Computer Engineering \\
University of California, San Diego \\
\texttt{\{kuc029, chl438, brao, hgarudadri\}@ucsd.edu}}}
}

\maketitle
\copyrightnotice
\begin{abstract}
We show that a new design criterion, i.e., the least squares on subband errors regularized by a weighted norm, can be used to generalize the proportionate-type normalized subband adaptive filtering (PtNSAF) framework. The new criterion directly penalizes subband errors and includes a sparsity penalty term which is minimized using the damped regularized Newton's method.
The impact of the proposed generalized PtNSAF (GPtNSAF) is studied for the system identification problem via computer simulations.
Specifically, we study the effects of using different numbers of subbands and various sparsity penalty terms for quasi-sparse, sparse, and dispersive systems.
The results show that the benefit of increasing the
number of subbands is larger than promoting sparsity of the estimated filter coefficients when the target system is quasi-sparse or dispersive.
On the other hand, for sparse target systems, promoting sparsity becomes more important.
More importantly, the two aspects provide complementary and additive benefits to the GPtNSAF for speeding up convergence.
\end{abstract}


\input{introduction}
\input{proposed_gpnsaf}
\input{simulation_results}
\input{conclusion}

\section*{Acknowledgment}
\footnotesize
This work was supported by NIH/NIDCD grants R01DC015436 and R33DC015046.

\bibliographystyle{IEEEtran}
\bibliography{ref}

\end{document}

%% file: introduction.tex
\section{Introduction}
The classic least mean square (LMS) and normalized LMS (NLMS) \cite{Widrow1985ASP,haykin2008adaptive,sayed2011adaptive} both show degraded convergence behaviors when the input signal is colored.
This problem can be addressed by whitening the colored input using a family of conventional subband adaptive filters (SAFs) \cite{lee2009subband} where each subband utilizes an adaptive filter independently. However, they are known to suffer from the problem of aliasing and band-edge effects \cite{lee2009subband}. To address this issue, a family of new SAFs has been proposed in \cite{courville1998weighted,reddy1999saf,lee2004subband} where each subband error signal is normalized by the corresponding input power and aggregated to update the fullband filter taps. It has been shown that the family of new SAFs can be derived from three different perspectives: i) gradient descent on weighted subband errors \cite{courville1998weighted}, ii) polyphase decomposition \cite{reddy1999saf}, and iii) constrained subband updates \cite{lee2004subband}. These new SAFs are termed normalized SAF (NSAF) due to their identical behavior. Hence, the NSAF can be viewed as a subband generalization of the NLMS \cite{lee2006leastperturbation}.

In \cite{duttweiler2000proportionate}, the proportionate NLMS (PNLMS) was introduced to improve convergence behavior by intuitively assigning a step size proportional to the magnitude of the estimated coefficient to each filter tap. Unfortunately, PNLMS tends to slow down after initial fast convergence \cite{benesty2002ipnlms}. Many PNLMS variants were later proposed to address this convergence issue and \cite{wagner2013proportionate} provides a good review.
Among those variants, the pNLMS \cite{rao2003slms} has been proposed as a generalization of PNLMS and was derived by minimizing a modified mean squared error criterion regularized by the \textit{p}-norm-like diversity measure. The $p$ value can be chosen to promote different degrees of sparsity and the effectiveness has been verified in the application of adaptive feedback cancellation \cite{lee2017slms}.

A family of proportionate NSAFs (PNSAFs) \cite{abadi2009proportionate,abadi2011family,see2008psaf,pradhan2017safha} has been proposed on top of NSAF to speed up the convergence of adaptive filters by simultaneously exploiting the sparse structure of the fullband filter taps and decorrelating the colored input signals. However, these PtNSAFs were proposed in an intuitive way and no theoretical convergence analysis was conducted. In \cite{puhan2019zero}, a zero-attracting PNSAF (ZA-PNSAF) was derived from an optimization criterion, yet, the proportionate matrix used does not have theoretical support and the ability to fit in with different degrees of sparsity. Besides, all these previous works use a decimation factor which is equal to the number of subbands in PNSAFs. Due to the nature of the proportionate matrix (function of current filter taps), an analysis with no decimation on PtNSAF is needed.

In this paper, we propose a generalized PtNSAF (GPtNSAF) which is  derived utilizing a well posed optimization criterion reflecting the filtering objectives,
as well as based on well founded optimization algorithmic principles. Furthermore, the proposed filtering structure can be operated on any decimation factor. We show that GPtNSAF is a generalization of the PtNSAF, proportionate-type affine projection algorithm (PtAPA), NSAF, PtNLMS, and NLMS. The effectiveness of the proposed adaptive filter is verified on different environments including quasi-sparse (compressible), sparse, and dispersive target systems via computer simulations.

\textit{Signal Model:}
\begin{figure}[ht!]
    \centering
    \includegraphics[width=0.5\textwidth]{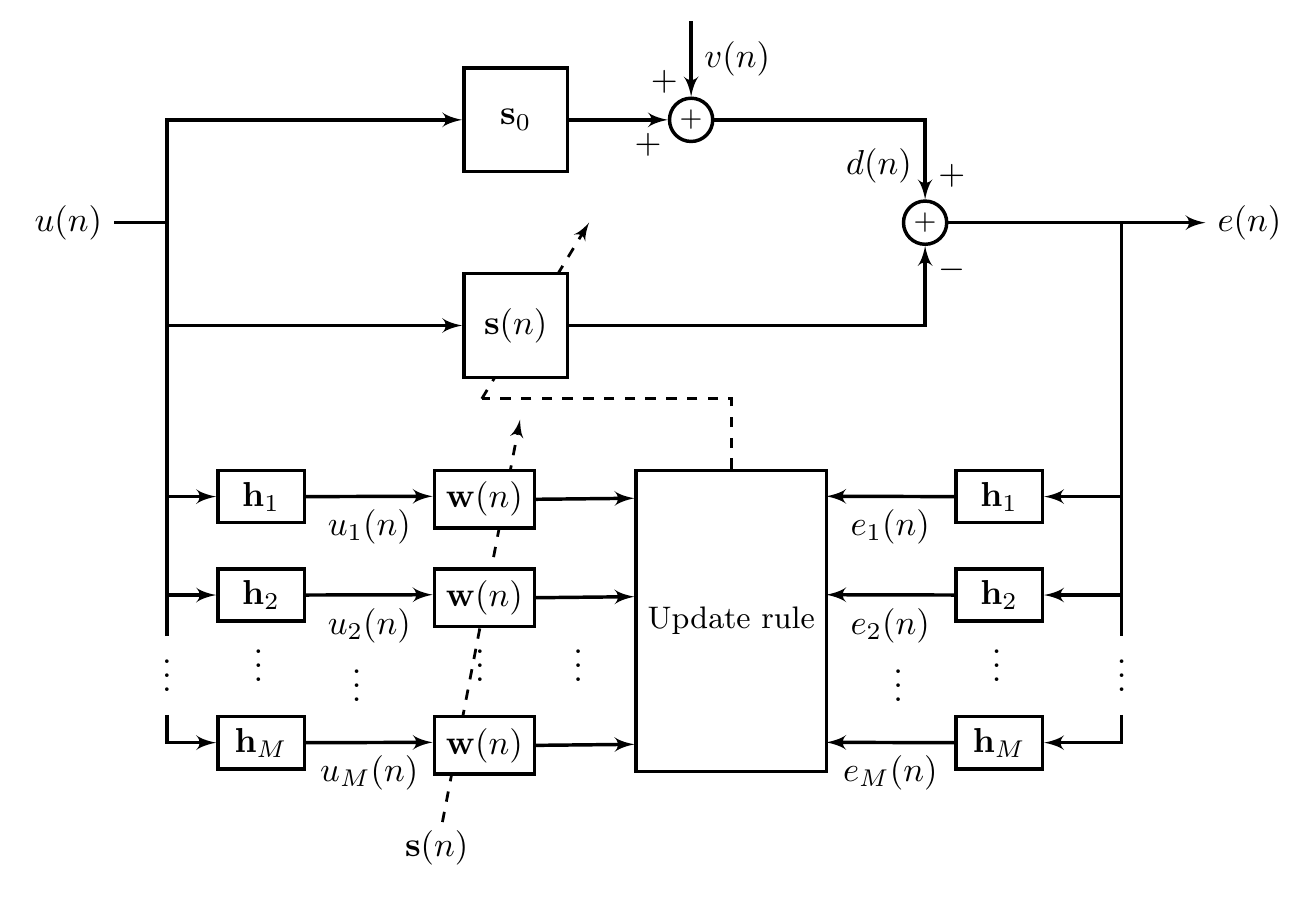}
    \caption{Block diagram of the GPtNSAF.}
    \label{fig:system_identification}
\end{figure}
Before deriving the proposed adaptive filter, we define some useful notations and the signal model in Fig. \ref{fig:system_identification}.
$\mathbf{H}\in\mathbb{R}^{N\times M}$ is an \textit{M}-channel analysis filter bank matrix where each of the analysis filter is of length $N$ and each column of $\mathbf{H}$ is a band-pass filter, i.e.,
$
    \mathbf{H}=\begin{bmatrix}\mathbf{h}_1&\mathbf{h}_2&\cdots&\mathbf{h}_M\end{bmatrix}.
$
$
    \mathbf{e}(n)=\begin{bmatrix}e(n)&e(n-1)&\cdots&e(n-N+1)\end{bmatrix}^T\in\mathbb{R}^N
$ is the fullband error vector
where $e(n)$ is the error in the fullband at time $n$.
$
    \mathbf{U}(n)=\begin{bmatrix}\mathbf{u}(n)&\mathbf{u}(n-1)&\cdots&\mathbf{u}(n-N+1)\end{bmatrix}\in\mathbb{R}^{L\times N}
$ is the fullband input data matrix
where $\mathbf{u}(n)=\begin{bmatrix}u(n)&u(n-1)&\cdots&u(n-L+1)\end{bmatrix}^T\in\mathbb{R}^L$ is the fullband input vector.
$\mathbf{e}_b(n)=\begin{bmatrix}e_1(n)&e_2(n)&\cdots&e_M(n)\end{bmatrix}^T=\begin{bmatrix}\mathbf{h}_1^T\mathbf{e}(n)&\mathbf{h}_2^T\mathbf{e}(n)&\cdots&\mathbf{h}_M^T\mathbf{e}(n)\end{bmatrix}^T=\mathbf{H}^T\mathbf{e}(n)\in\mathbb{R}^{M}$ is the subband error vector
where $e_i(n)\in\mathbb{R}$ is the error in the $i^{th}$ subband at time $n$.
$
    \mathbf{U}_b(n)=\begin{bmatrix}\mathbf{u}_1(n)&\mathbf{u}_2(n)&\cdots&\mathbf{u}_M(n)\end{bmatrix}=\mathbf{U}(n)\mathbf{H}\in\mathbb{R}^{L\times M}
$ is the subband input data matrix
where $\mathbf{u}_i(n)=\begin{bmatrix}u_i(n)&u_i(n-1)&\cdots&u_i(n-L+1)\end{bmatrix}^T\in\mathbb{R}^{L}$ is the input vector at the $i^{th}$ subband which can be computed by $\mathbf{u}_i(n)=\mathbf{U}(n)\mathbf{h}_i$.
Typically, we require $L\geq M$ to avoid the overcomplete representation of the input signal and the singularity introduced by the correlation matrix of the subband input data matrix \cite{lee2004subband}.
Next, by defining the fullband desire vector as
$
    \mathbf{d}(n)=\begin{bmatrix}d(n)&d(n-1)&\cdots&d(n-N+1)\end{bmatrix}^T\in\mathbb{R}^N
$ where $d(n)=\mathbf{u}^T(n)\mathbf{s}_0+v(n)$,
we are able to expand the fullband error vector as
$
        \mathbf{e}(n)
        =\mathbf{d}(n)-\mathbf{U}^T(n)\mathbf{s}(n).
$
$\mathbf{s}_0\in\mathbb{R}^L$ denotes the target system and $v(n)\in\mathbb{R}$ is the system noise.
$\mathbf{s}(n)=\begin{bmatrix}s_1(n)&s_2(n)&\cdots&s_L(n)\end{bmatrix}^T\in\mathbb{R}^L$ is the adaptive filter taps at time $n$.
Finally, we define the proportionate matrix $\mathbf{W}(n)=\text{diag}\{w_1(n),w_2(n),\cdots,w_{L}(n)\}\in\mathbb{R}^{L\times L}$ as a positive definite matrix which promotes the sparse structure of $\mathbf{s}(n+1)$; it takes the adaptive filter $\mathbf{s}(n)$ at current time $n$ as its input; hence, $\mathbf{W}(n)$ is given in each iteration. In Fig. \ref{fig:system_identification}, $\mathbf{w}(n)\in\mathbb{R}^L$ is defined as $\mathbf{w}(n)=\begin{bmatrix}\sqrt{w_1(n)}&\sqrt{w_2(n)}&\cdots&\sqrt{w_L(n)}\end{bmatrix}^T$. Finally, $\mathbf{s}\in\mathbb{R}^{L}$ is the adaptive filter which is an optimization variable.

%% file: proposed_gpnsaf.tex
\section{The Proposed Generalized Proportionate-Type Normalized Subband Adaptive Filter}
In this section, we propose a novel optimization criterion and the derivation for the GPtNSAF which exploits the structures of the input signal and the underlying unknown (target) system. We find that the PtNSAF, PtAPA, NSAF, PtNLMS, and NLMS are all special cases and can be obtained by using different settings of the hyperparameters in GPtNSAF. We focus on the derivation where there is no decimation. One can easily show that the proposed adaptive filter can be readily extended to include an arbitrary decimation factor.

\subsection{The Proposed Criterion: The Least Squares on Subband Errors Regularized by a Weighted Norm}
Instead of minimizing the fullband squared error \cite{lee2019slms}, we minimize the sum of the squared error in each subband with a sparsity penalty term. We propose the following cost function:
\begin{equation} \label{eq:opt_problem_1}
    J(\mathbf{s})\triangleq\sum_{i=1}^{M}\abs{e_i(n,\mathbf{s})}^2+\tau\norm{\mathbf{s}}_{\mathbf{W}^{-1}(n)}^2
\end{equation}
where
$
    e_i(n,\mathbf{s})\triangleq\mathbf{h}_i^T\mathbf{e}(n,\mathbf{s})=\mathbf{h}_i^T\left[\mathbf{d}(n)-\mathbf{U}^T(n)\mathbf{s}\right]
$
and we have used $\norm{\mathbf{s}}_{\mathbf{W}^{-1}(n)}^2$ to stand for the weighted norm squared $\mathbf{s}^T\mathbf{W}^{-1}(n)\mathbf{s}$; this regularization term is designed to expedite the system identification process by introducing a weighted norm for filter taps.
In this paper, we use the $\mathbf{W}(n)$ suggested in \cite{rao2003slms,lee2017slms,lee2019slms} for promoting different degrees of sparsity due to its theoretical support.
Since the task of correctly identifying the underlying unknown system is more important than promoting the sparsity of the filter taps, the regularization parameter $\tau>0$ is set to a very small number. In order to find an LMS-like adaptation, we minimize the cost function (\ref{eq:opt_problem_1}) by using the damped regularized Newton's method.

\subsection{Deriving GPtNSAF}
To proceed, we perform the affine scaling transform (AST) \cite{rao1999ast} on the optimization variable $\mathbf{s}$:
\begin{equation} \label{eq:ast}
    \mathbf{q}\triangleq\mathbf{W}^{-\frac{1}{2}}(n)\mathbf{s}.
\end{equation}
Applying (\ref{eq:ast}) into (\ref{eq:opt_problem_1}) , the equivalent optimization problem $
    \min_{\mathbf{q}} J(\mathbf{q})=\sum_{i=1}^{M}\abs{e_i(n,\mathbf{W}^{\frac{1}{2}}(n)\mathbf{q})}^2+\tau\norm{\mathbf{q}}_2^2
$ in $\mathbf{q}$ domain can be easily solved.
We define the \textit{a posteriori} AST variable at time $n$ as $\mathbf{q}(n|n)\triangleq\mathbf{W}^{-\frac{1}{2}}(n)\mathbf{s}(n)$ and the \textit{a priori} AST variable as $\mathbf{q}(n+1|n)\triangleq\mathbf{W}^{-\frac{1}{2}}(n)\mathbf{s}(n+1)$.

Now, we consider the damped regularized Newton's method for the update rule on minimizing $J(\mathbf{q})$, i.e., 
$
    \mathbf{q}(n+1|n) =\mathbf{q}(n|n) - \mu \left[\nabla_{\mathbf{\mathbf{q}}}^2J\left(\mathbf{q}(n|n)\right)+2\delta\mathbf{I}\right]^{-1}\nabla_{\mathbf{\mathbf{q}}} J\left(\mathbf{q}(n|n)\right)
$ where $\mu>0$ is the step size for adaptation and $\delta>0$ is a regularization parameter.
The gradient of $J(\mathbf{q})$ is given by
\begin{equation}
        \nabla_{\mathbf{q}}J(\mathbf{q}(n|n))
        =-2\mathbf{W}^{\frac{1}{2}}(n)\mathbf{U}_b(n)\mathbf{e}_b(n)+2\tau\mathbf{q}(n|n).
\end{equation}
Next, the Hessian is given by
\begin{equation}
        \nabla_{\mathbf{q}}^2J(\mathbf{q}(n|n))
        =2\mathbf{W}^{\frac{1}{2}}(n)\mathbf{U}_b(n)\mathbf{U}_b^T(n)\mathbf{W}^{\frac{1}{2}}(n)+2\tau\mathbf{I}.
\end{equation}
Therefore, the update rule on $\mathbf{q}$ domain is given by
\begin{equation} \label{eq:update_rule_q}
    \begin{split}
        \mathbf{q}(n+1|n)&=\left(\mathbf{I}-\frac{\mu\tau}{\delta+\tau}\left[\mathbf{I}-\boldsymbol{\Psi}(n)\right]\right)\mathbf{q}(n|n)\\&\ \ \ \ +\mu\mathbf{W}^{\frac{1}{2}}(n)\mathbf{U}_b(n)\boldsymbol{\Phi}(n)\mathbf{e}_b(n)
    \end{split}
\end{equation}
where we have applied the Woodbury matrix identity to avoid large matrix inversion (\textit{L}-by-\textit{L}) in the damped regularized Newton's method and
\begin{equation}
    \boldsymbol{\Psi}(n)\triangleq\mathbf{W}^{\frac{1}{2}}(n)\mathbf{U}_b(n)\boldsymbol{\Phi}(n)\mathbf{U}_b^T(n)\mathbf{W}^{\frac{1}{2}}(n).
\end{equation}
Notice that the inverse of the regularized weighted subband correlation matrix, i.e.,
\begin{equation}
    \boldsymbol{\Phi}(n)\triangleq\left[(\delta+\tau)\mathbf{I}_M+\mathbf{U}_b^T(n)\mathbf{W}(n)\mathbf{U}_b(n)\right]^{-1}
\end{equation}
is an \textit{M}-by-\textit{M} matrix inversion ($L \gg M$ in most cases).
Converting $\mathbf{q}$ back to the $\mathbf{s}$ domain, we have
\begin{equation} \label{eq:update_rule_s}
    \begin{split}
        \mathbf{s}(n+1)&=\left(\mathbf{I}-\frac{\mu\tau}{\delta+\tau}\left[\mathbf{I}-\boldsymbol{\Psi}(n)\right]\right)\mathbf{s}(n)\\&\ \ \ \ +\mu\mathbf{W}(n)\mathbf{U}_b(n)\boldsymbol{\Phi}(n)\mathbf{e}_b(n).
    \end{split}
\end{equation}
Finally, setting $\tau\to 0^+$ leads to the update rule for the GPtNSAF:
$
    \mathbf{s}(n+1)=\mathbf{s}(n)+\mu\mathbf{g}(n)
$
where
\begin{equation}
    \mathbf{g}(n)=\mathbf{W}(n)\mathbf{U}_b(n)\left[\delta\mathbf{I}_M+\mathbf{U}_b^T(n)\mathbf{W}(n)\mathbf{U}_b(n)\right]^{-1}\mathbf{e}_b(n).
\end{equation}
\subsection{Special Cases of the GPtNSAF}
\subsubsection{PtNSAF}
By selecting $\mathbf{H}$ as the set of eigenvectors for the weighted correlation matrix $\mathbf{U}^T(n)\mathbf{W}(n)\mathbf{U}(n)$, we have the PtNSAF:
$
    \mathbf{g}(n)=\sum_{i=1}^M\frac{e_i(n)}{\mathbf{u}_i^T(n)\mathbf{W}(n)\mathbf{u}_i(n)+\delta}\mathbf{W}(n)\mathbf{u}_i(n).
$
\subsubsection{PtAPA}
By choosing $\mathbf{H}=\mathbf{I}$, we have the PtAPA:
$
    \mathbf{g}(n)=\mathbf{W}(n)\mathbf{U}(n)\left[\delta\mathbf{I}_M+\mathbf{U}^T(n)\mathbf{W}(n)\mathbf{U}(n)\right]^{-1}\mathbf{e}(n).
$ Obviously, the APA is directly followed by setting $\mathbf{W}(n)=\mathbf{I}$.
\subsubsection{NSAF}
Based on PtNSAF, setting $\mathbf{W}(n)=\mathbf{I}$ gives the NSAF:
$
    \mathbf{g}(n)=\sum_{i=1}^M\frac{e_i(n)}{\mathbf{u}_i^T(n)\mathbf{u}_i(n)+\delta}\mathbf{u}_i(n).
$
\subsubsection{PtNLMS}
Setting $M=N=1$ yields $\mathbf{H}=1\in\mathbb{R}$, thus we get the PtNLMS:
$
    \mathbf{g}(n)=\frac{e(n)}{\mathbf{u}^T(n)\mathbf{W}(n)\mathbf{u}(n)+\delta}\mathbf{W}(n)\mathbf{u}(n).
$
\subsubsection{NLMS}
Based on PtNLMS, setting $\mathbf{W}(n)=\mathbf{I}$ gives the NLMS:
$
    \mathbf{g}(n)=\frac{e(n)}{\mathbf{u}^T(n)\mathbf{u}(n)+\delta}\mathbf{u}(n).
$

%% file: simulation_results.tex
\section{Simulation Results}
\begin{figure*}
    \centering
    \begin{subfigure}[b]{0.32\textwidth}
        \includegraphics[width=0.9\textwidth]{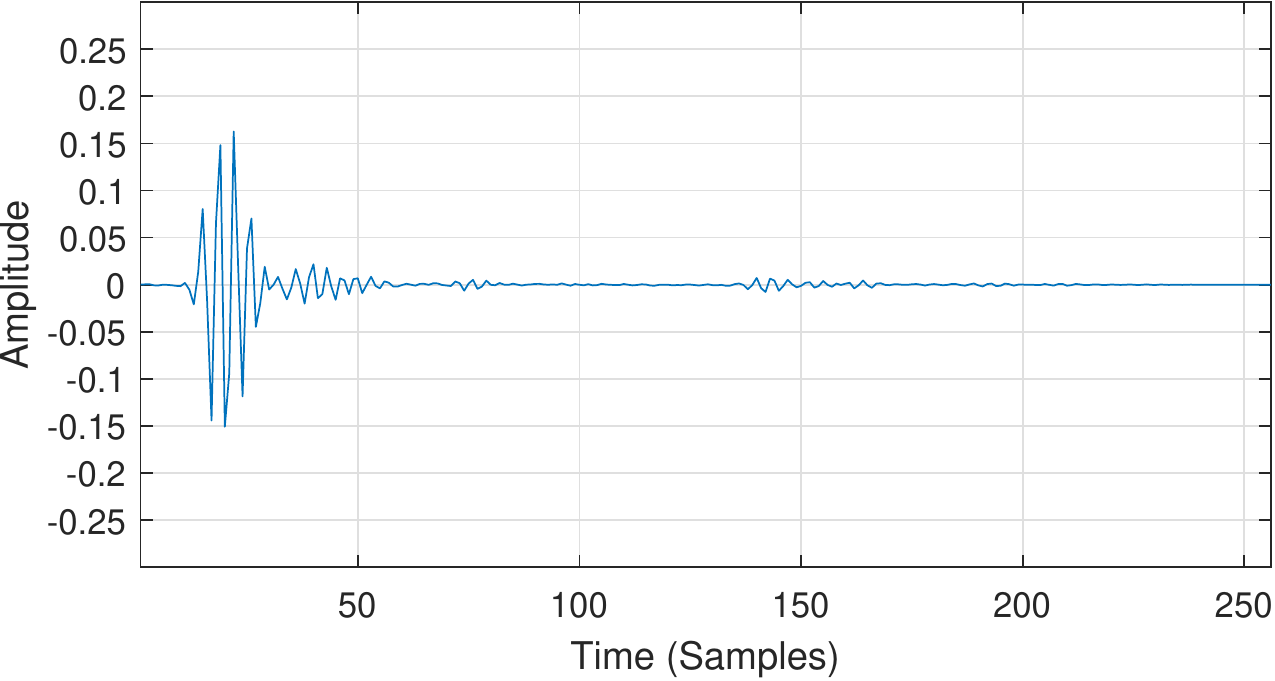}
        \caption{A quasi-sparse IR.}
        \label{fig:quasi_sparse_ir}
    \end{subfigure}
    \begin{subfigure}[b]{0.32\textwidth}
        \includegraphics[width=0.9\textwidth]{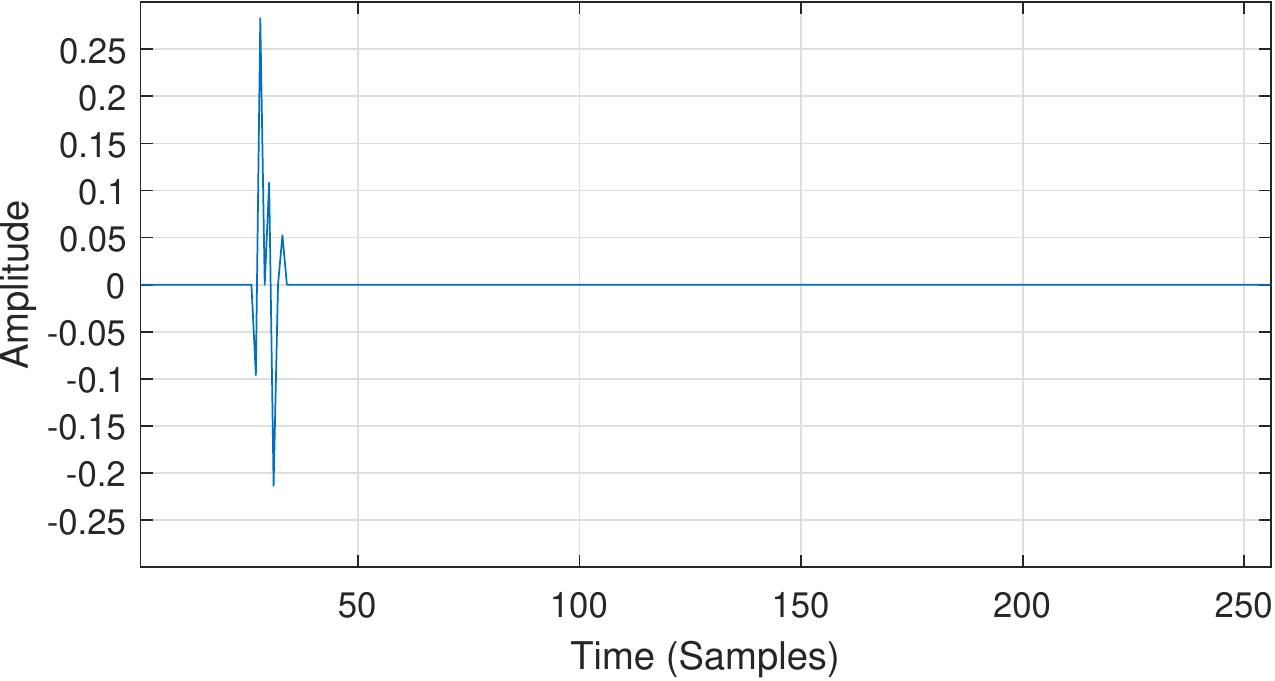}
        \caption{A sparse IR.}
        \label{fig:sparse_ir}
    \end{subfigure}
    \begin{subfigure}[b]{0.32\textwidth}
        \includegraphics[width=0.9\textwidth]{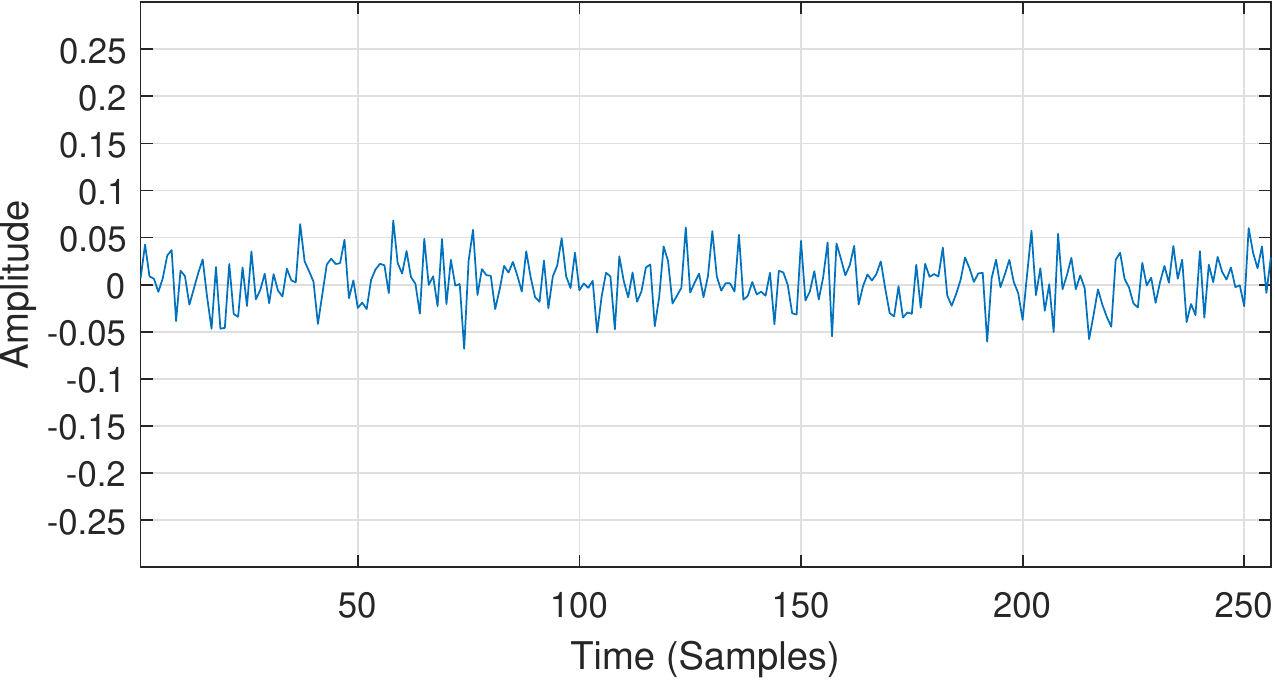}
        \caption{A dispersive IR.}
        \label{fig:dispersive_ir}
    \end{subfigure}
    \caption{(a), (b), and (c) are three different IRs (target systems) of length $L=256$ with different degrees of sparsity. (a) is a measured acoustic feedback path IR. (b) and (c) are artificial IRs. Notice that the IRs in (a), (b), and (c) have the same energy.}
    \label{fig:three_ir}
\end{figure*}
We study the convergence performance of the proposed GPtNSAF and some of its special cases including PtNSAF, NSAF, PtNLMS, and NLMS in a system identification scenario via computer simulations.

\subsection{Experimental Setup}
The impulse responses (IRs) of the target systems are shown in Fig. \ref{fig:three_ir}. The input signal is a first order autoregressive (AR) process defined by
$
    u(n) = \rho u(n-1)+x(n)
$
where $\rho=0.9$ and $x(n)$ is a zero mean and unit variance white Gaussian noise. In the simulations, we discarded the first 2000 samples of $u(n)$ to make sure the stationarity of the AR process. The system noise $v(n)$ is a zero mean white Gaussian noise with variance $\sigma_v^2=10^{-3}$ which gives $-30$ dB noise level ($J_{min}$). The length of the adaptive filter $L=256$ was set to the same size as in Fig. \ref{fig:three_ir} and all taps were initialized by $0$.

The analysis bank $\mathbf{H}$ is a cosine-modulated pseudo-quadrature mirror filter
(QMF) bank. We maintain the same transition bandwidth of the analysis filters for $M=2,4,8$ so that the comparison is fair. Therefore, the length of the analysis filter $N$ goes up with the number of subbands $M$. For $M=1,2,4,8$, we use $N=1,16,30,60$, respectively.

We used the sparsity promoting proportionate matrix
\begin{equation}
    w_i(n)=\left(\abs{s_i(n)}+c\right)^{2-p}, i=1,2,\cdots,L
\end{equation}
suggested in \cite{rao2003slms,lee2017slms,lee2019slms} so that we were allowed to adjust the degree of promoting sparsity by a single scalar $p\in[1.0,2.0]$ and a regularization parameter $c>0$.

The mean squared error (MSE) at time $n$ is defined by
$
    J(n)=\mathbb{E}\left[\abs{e(n)}^2\right]
$ where $\mathbb{E}[\cdot]$ denotes the mathematical expectation.
The MSE curves were obtained as the ensemble average over $1000$ Monte Carlo runs and normalized to start from $0$ dB. For all MSE simulations, we used $\mu=\frac{0.2}{M}$, $\delta=10^{-6}$, and $c=10^{-3}$.

\subsection{Studying $M$ and $\mathbf{W}(n)$ in GPtNSAF}
\begin{figure*}
    \centering
    \begin{subfigure}[b]{0.32\textwidth}
        \includegraphics[width=0.9\textwidth]{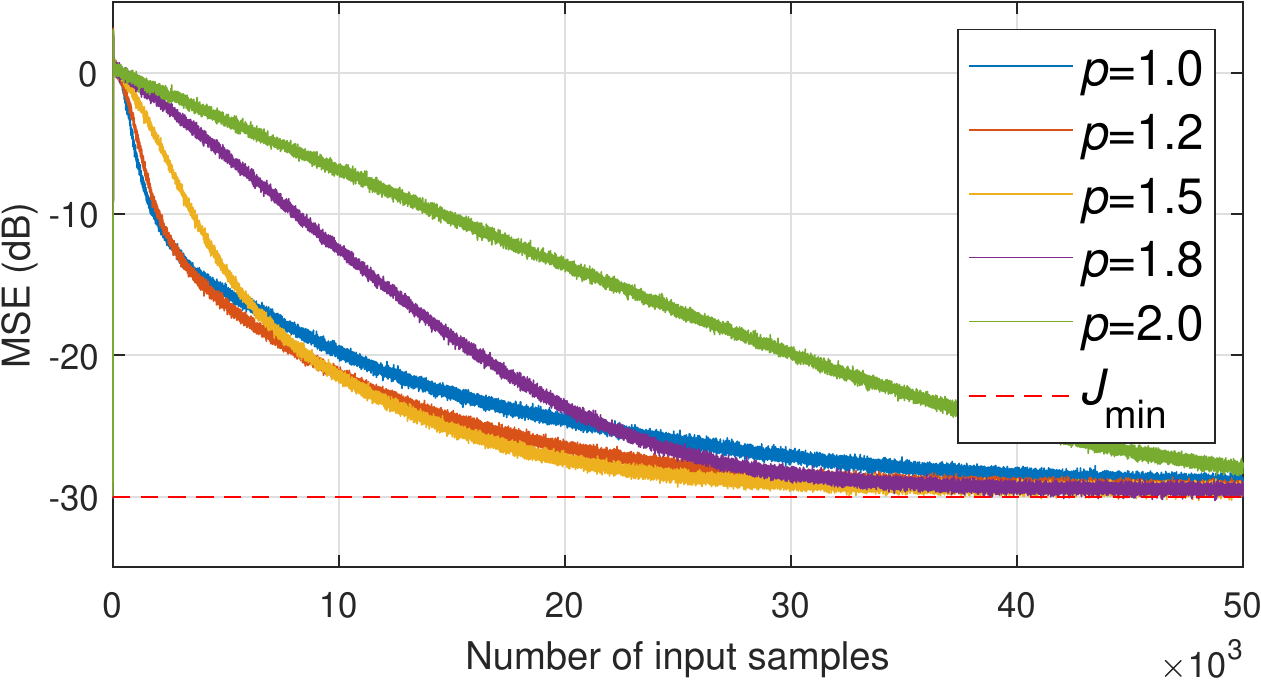}
        \caption{$M=1$}
        \label{fig:quasi_sparse_1}
        \vspace{0.6cm}
    \end{subfigure}
    \begin{subfigure}[b]{0.32\textwidth}
        \includegraphics[width=0.9\textwidth]{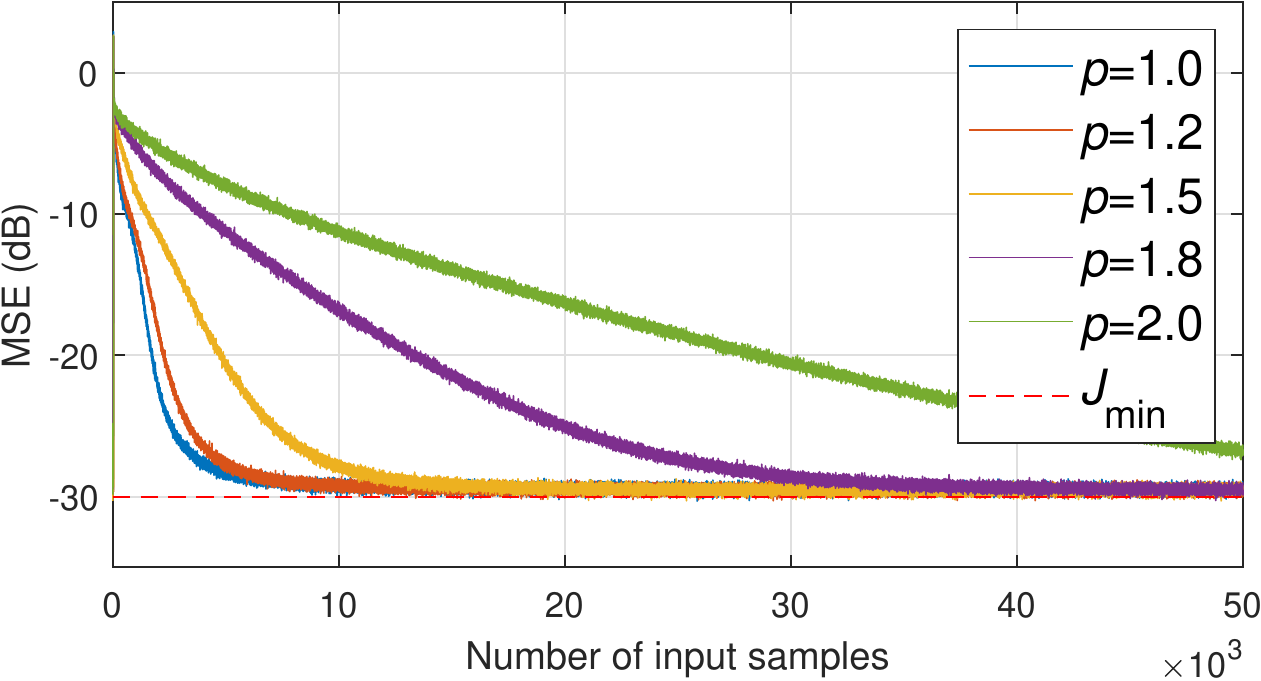}
        \caption{$M=1$}
        \label{fig:sparse_1}
        \vspace{0.6cm}
    \end{subfigure}
    \begin{subfigure}[b]{0.32\textwidth}
        \includegraphics[width=0.9\textwidth]{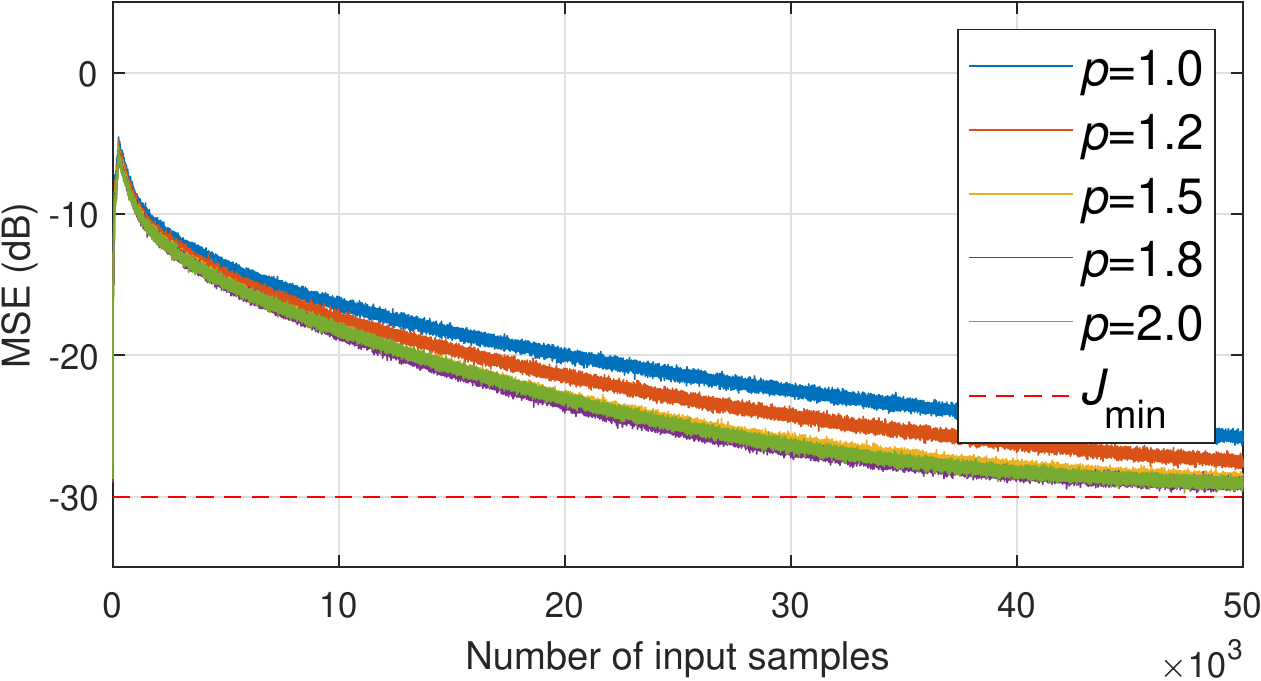}
        \caption{$M=1$}
        \label{fig:dispersive_1}
        \vspace{0.6cm}
    \end{subfigure}
    \begin{subfigure}[b]{0.32\textwidth}
        \includegraphics[width=0.9\textwidth]{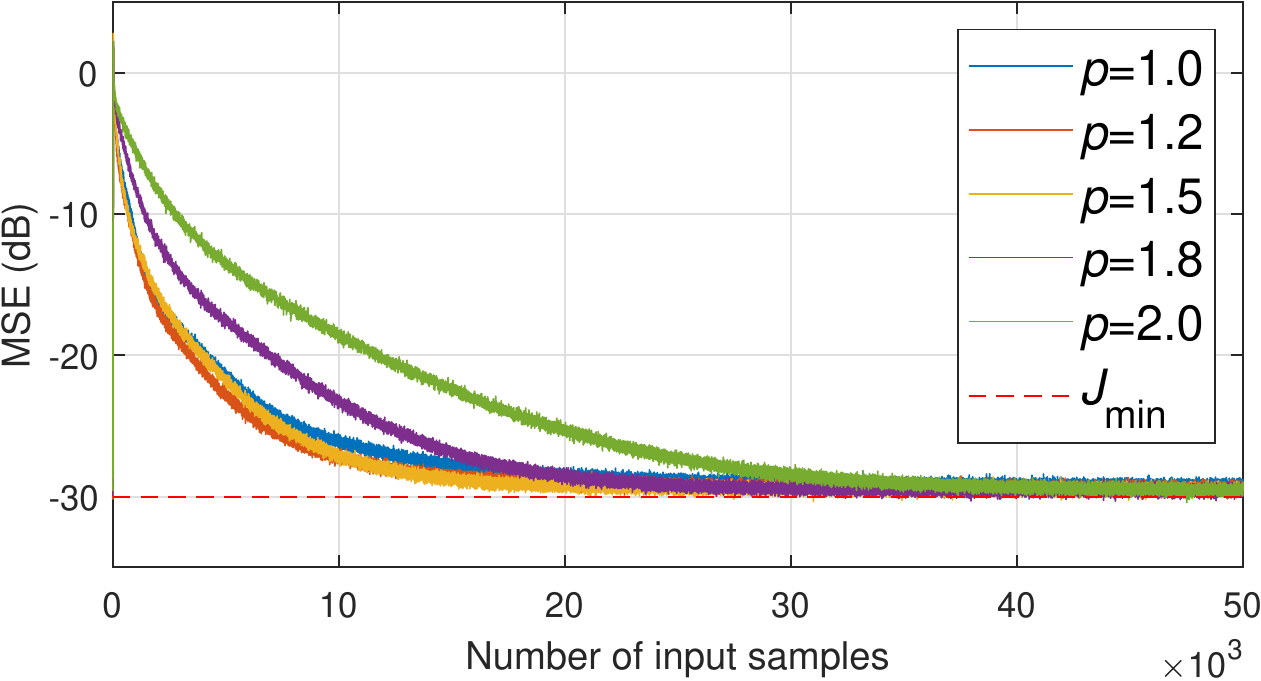}
        \caption{$M=2$}
        \label{fig:quasi_sparse_2}
        \vspace{0.6cm}
    \end{subfigure}
    \begin{subfigure}[b]{0.32\textwidth}
        \includegraphics[width=0.9\textwidth]{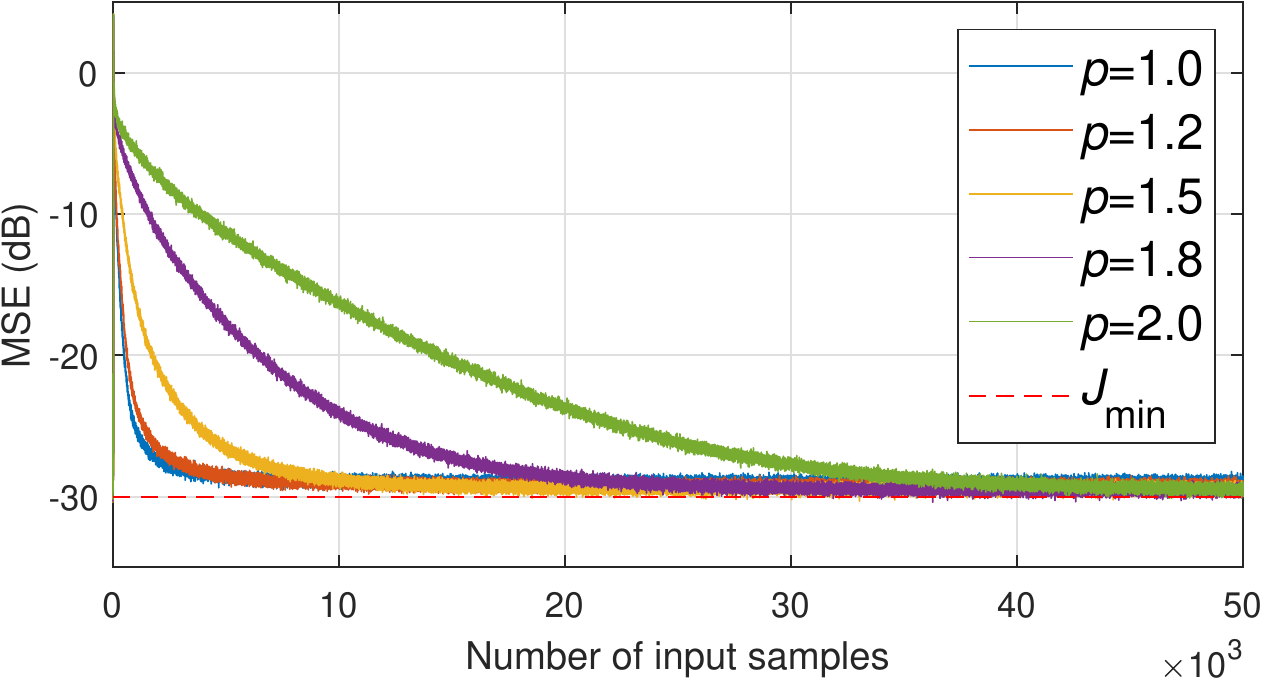}
        \caption{$M=2$}
        \label{fig:sparse_2}
        \vspace{0.6cm}
    \end{subfigure}
    \begin{subfigure}[b]{0.32\textwidth}
        \includegraphics[width=0.9\textwidth]{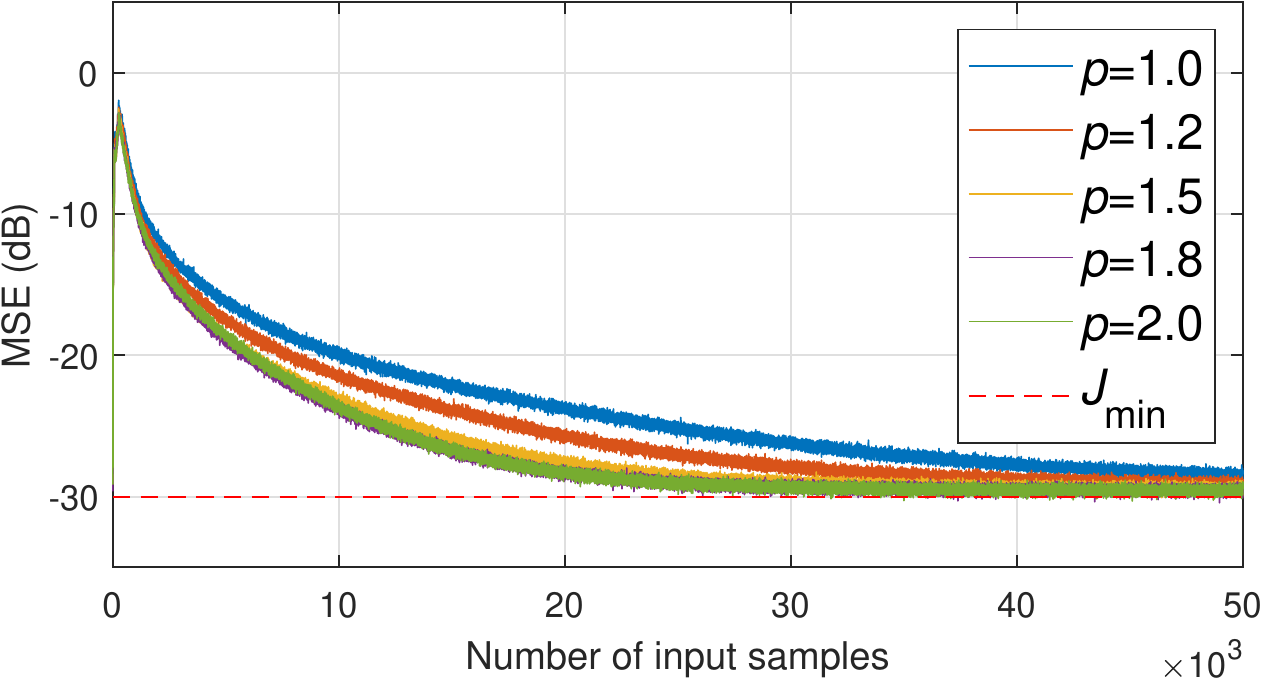}
        \caption{$M=2$}
        \label{fig:dispersive_2}
        \vspace{0.6cm}
    \end{subfigure}
    \begin{subfigure}[b]{0.32\textwidth}
        \includegraphics[width=0.9\textwidth]{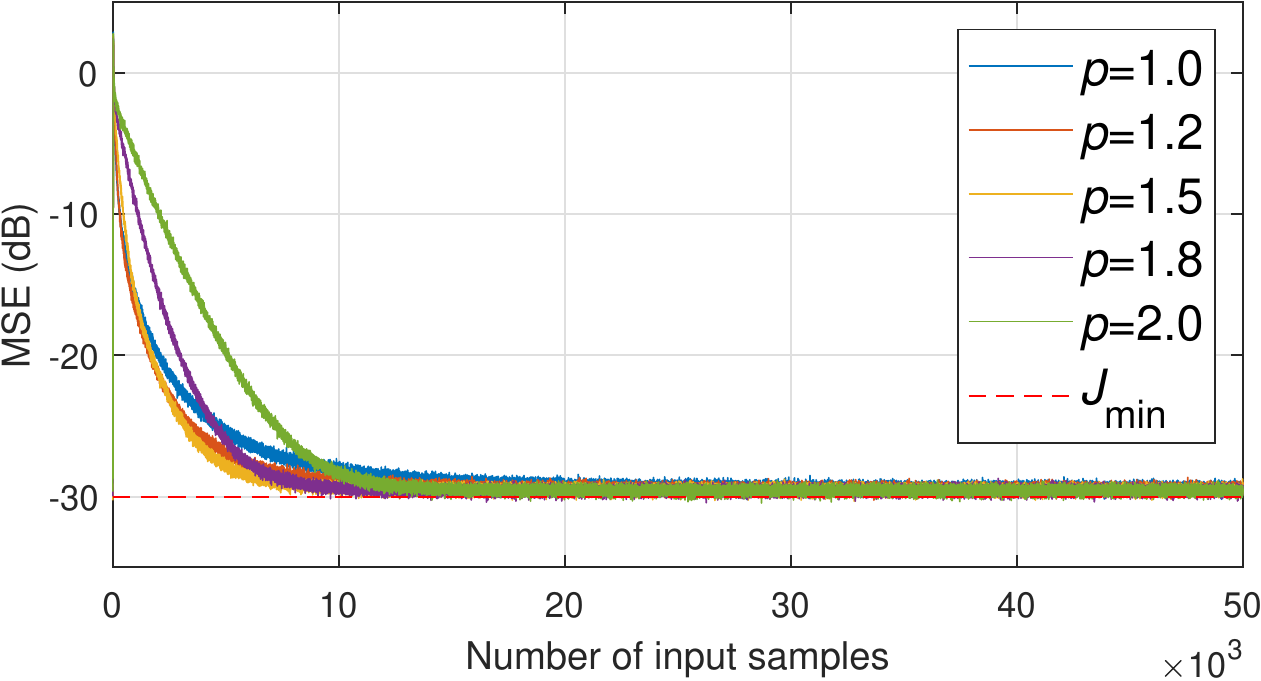}
        \caption{$M=4$}
        \label{fig:quasi_sparse_4}
        \vspace{0.6cm}
    \end{subfigure}
    \begin{subfigure}[b]{0.32\textwidth}
        \includegraphics[width=0.9\textwidth]{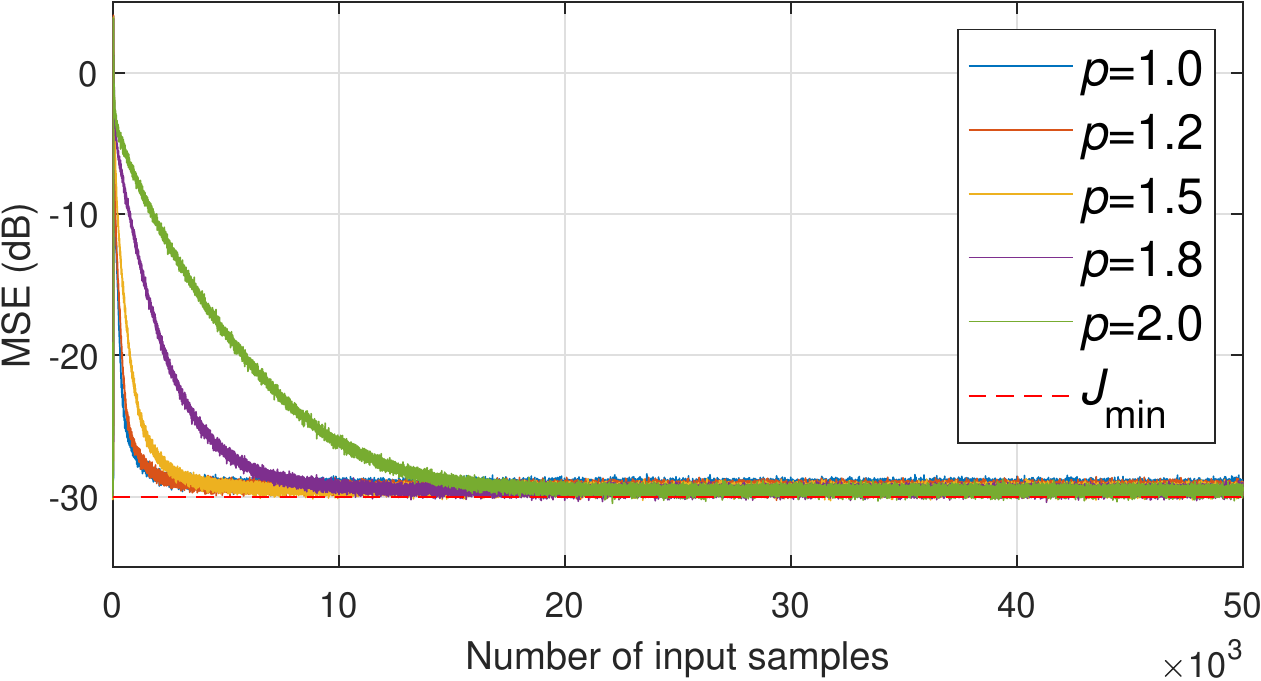}
        \caption{$M=4$}
        \label{fig:sparse_4}
        \vspace{0.6cm}
    \end{subfigure}
    \begin{subfigure}[b]{0.32\textwidth}
        \includegraphics[width=0.9\textwidth]{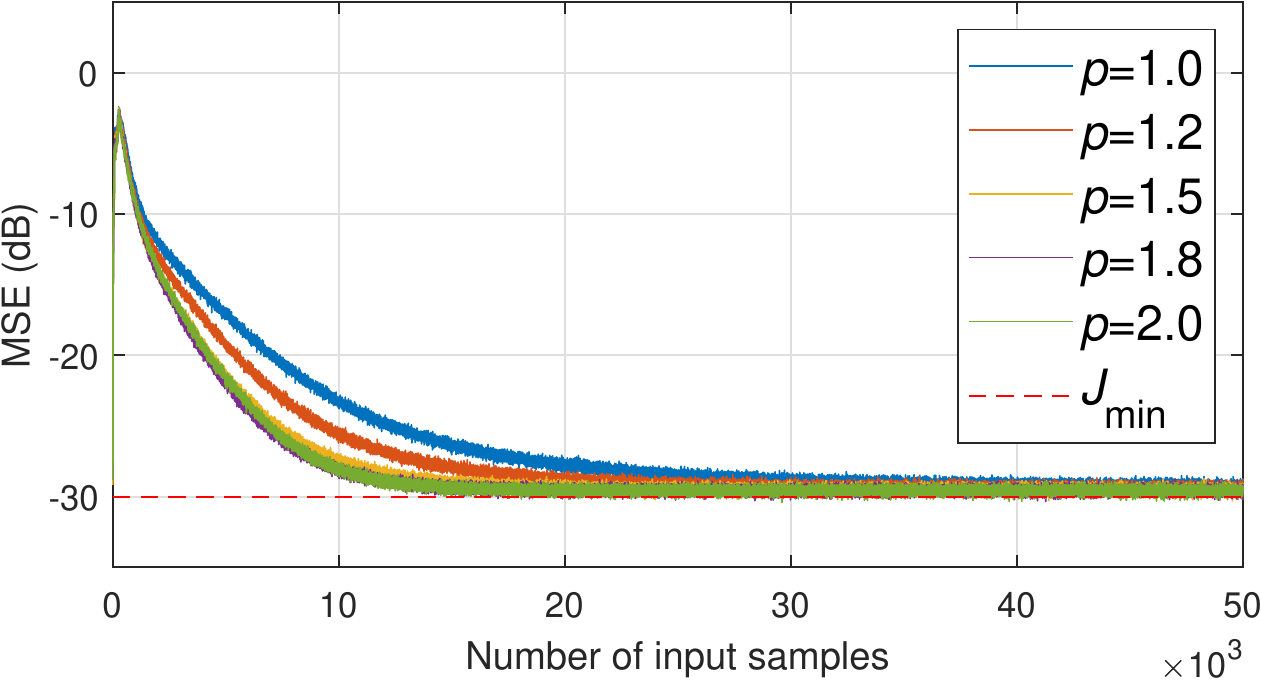}
        \caption{$M=4$}
        \label{fig:dispersive_4}
        \vspace{0.6cm}
    \end{subfigure}
    \begin{subfigure}[b]{0.32\textwidth}
        \includegraphics[width=0.9\textwidth]{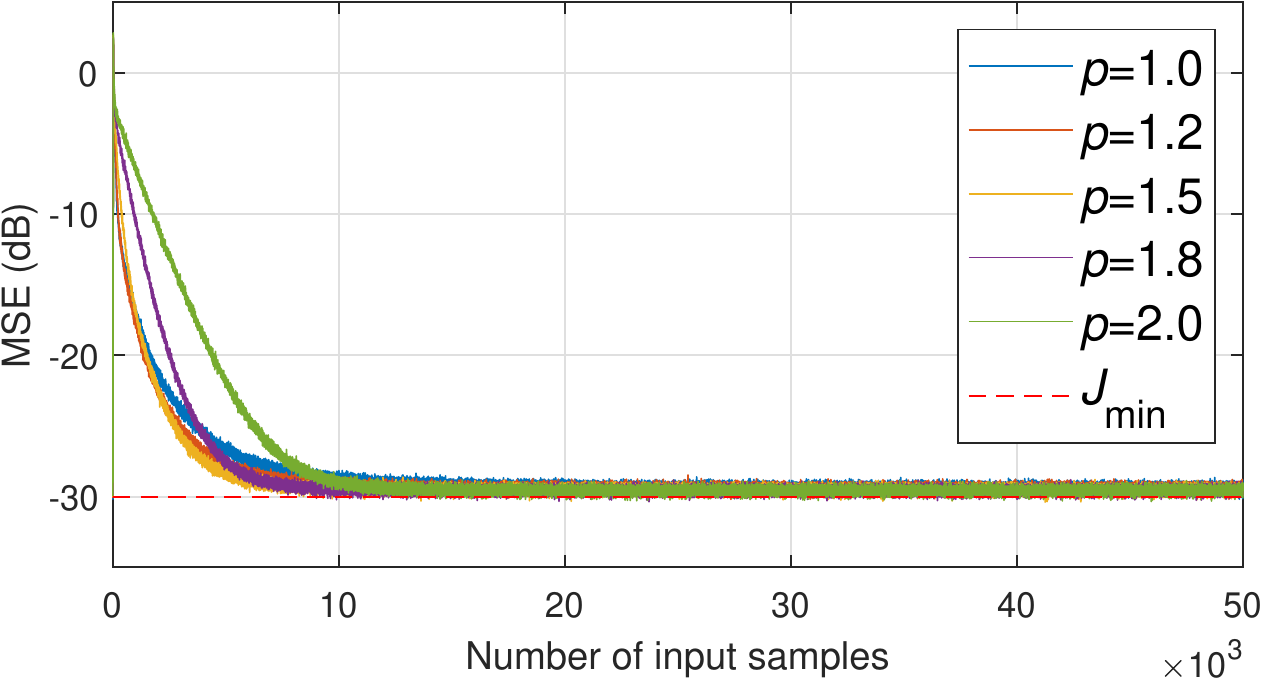}
        \caption{$M=8$}
        \label{fig:quasi_sparse_8}
    \end{subfigure}
    \begin{subfigure}[b]{0.32\textwidth}
        \includegraphics[width=0.9\textwidth]{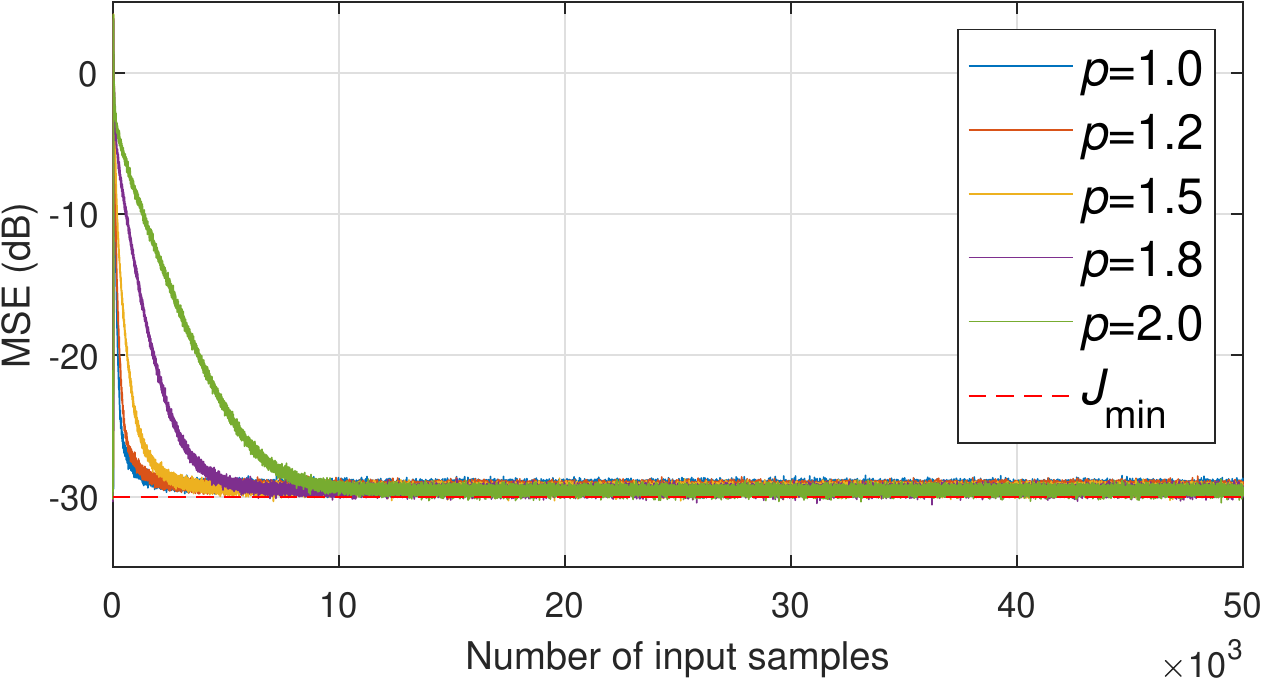}
        \caption{$M=8$}
        \label{fig:sparse_8}
    \end{subfigure}
    \begin{subfigure}[b]{0.32\textwidth}
        \includegraphics[width=0.9\textwidth]{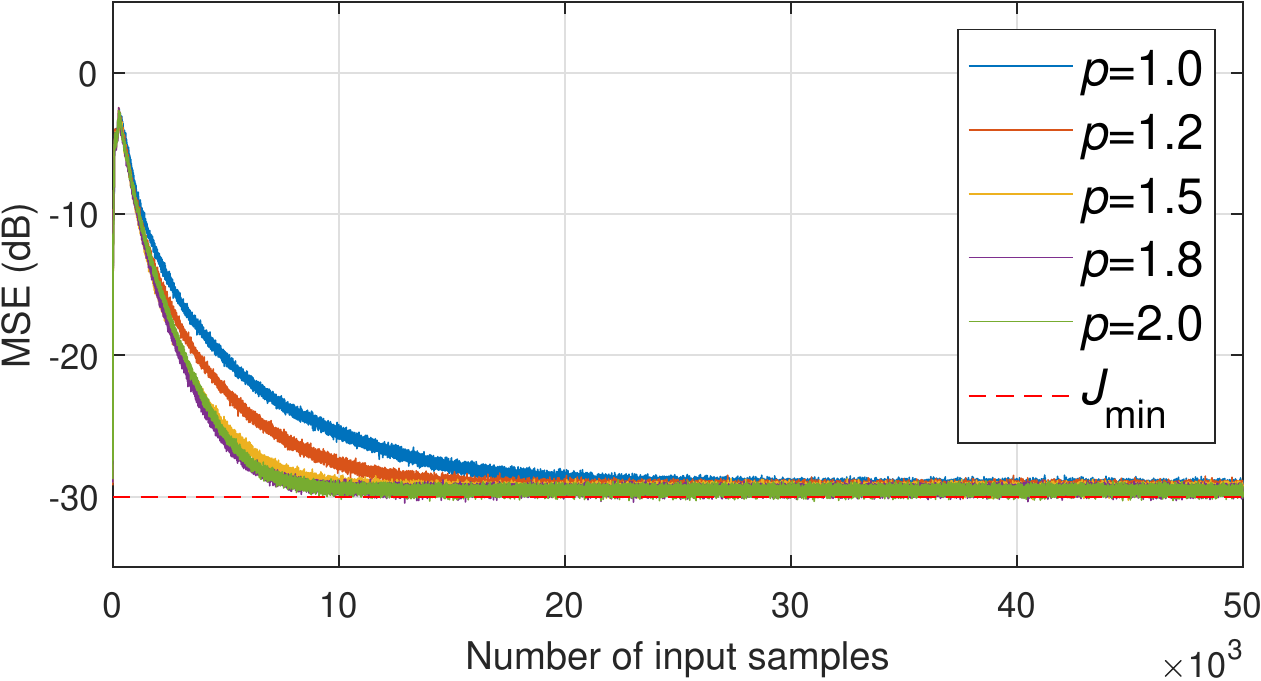}
        \caption{$M=8$}
        \label{fig:dispersive_8}
    \end{subfigure}
    \caption{The MSE curves of GPtNSAF using sparsity promoting proportionate matrix with different $p$ values for $M=1,2,4,8$. The target system for (a), (d), (g), and (j) is in Fig. \ref{fig:three_ir}(a); (b), (e), (h), and (k) is in Fig. \ref{fig:three_ir}(b); (c), (f), (i), and (l) is in Fig. \ref{fig:three_ir}(c). For the sake of comparing the different $\mathbf{W}(n)$, $M$, and target systems which have different degrees of sparsity, we visualize these MSE curves with the same number of input samples here.}
    \label{fig:12_mse}
\end{figure*}

We aim to show that the benefits of increasing the number of subbands $M$ and incorporating the proportionate matrix $\mathbf{W}(n)$ are complementary and additive for fast convergence. Fig. \ref{fig:12_mse} shows the MSE curves of GPtNSAF using sparsity promoting proportionate matrix with different $p$ values for $M=1,2,4,8$. According to the convergence behaviors in Fig. \ref{fig:12_mse}, the best $p$ values on each target system are consistent across different numbers of subbands. Therefore, we suggest $p\in[1.2,1.5]$, $p\in[1.0,1.2]$, and $p\in[1.8,2.0]$ for quasi-sparse, sparse, and dispersive target systems, respectively. Notice that the convergence speed is significantly improved for all target systems as the number of subbands increases. However, the performance gain is saturated at $M=8$. This is mainly due to the design of the analysis filter bank in which we did not emphasize on any particular bands. Instead, the spectrum is equally divided by the cosine-modulated pseudo-QMF bank.

\begin{figure*}
    \centering
    \begin{subfigure}[b]{0.32\textwidth}
        \includegraphics[width=0.9\textwidth]{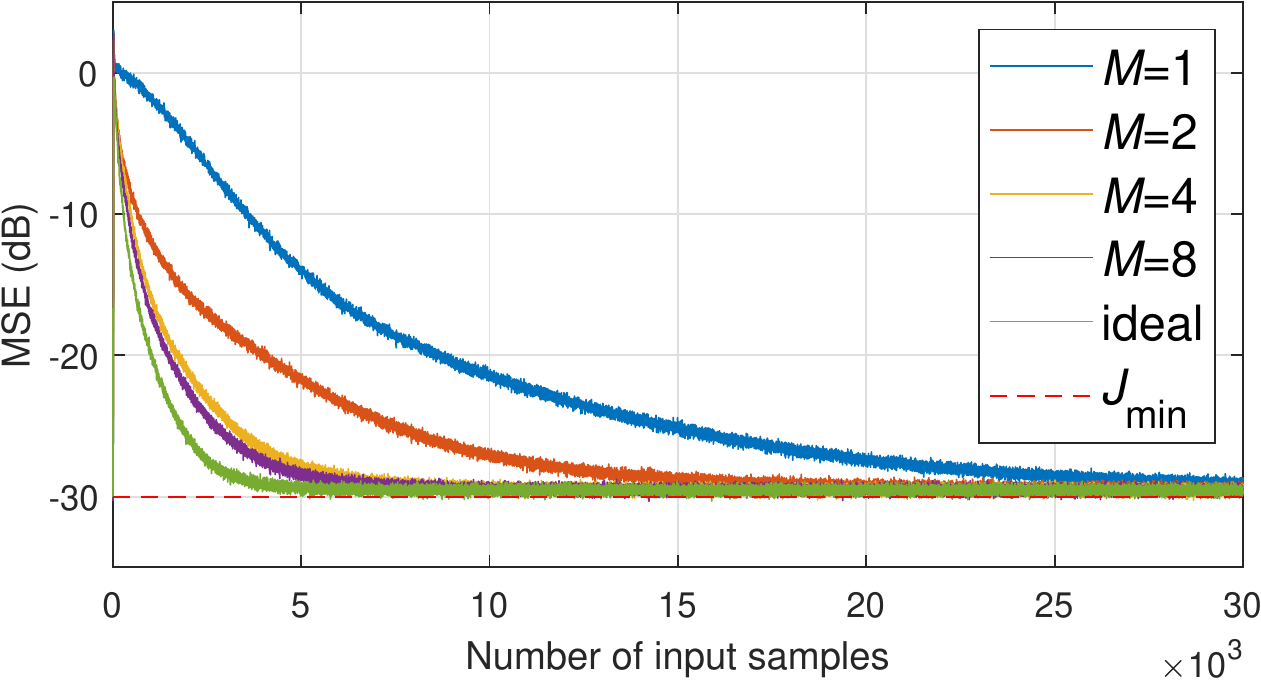}
        \caption{$p=1.5$ for the quasi-sparse target.}
        \label{fig:quasi_sparse_best}
    \end{subfigure}
    \begin{subfigure}[b]{0.32\textwidth}
        \includegraphics[width=0.9\textwidth]{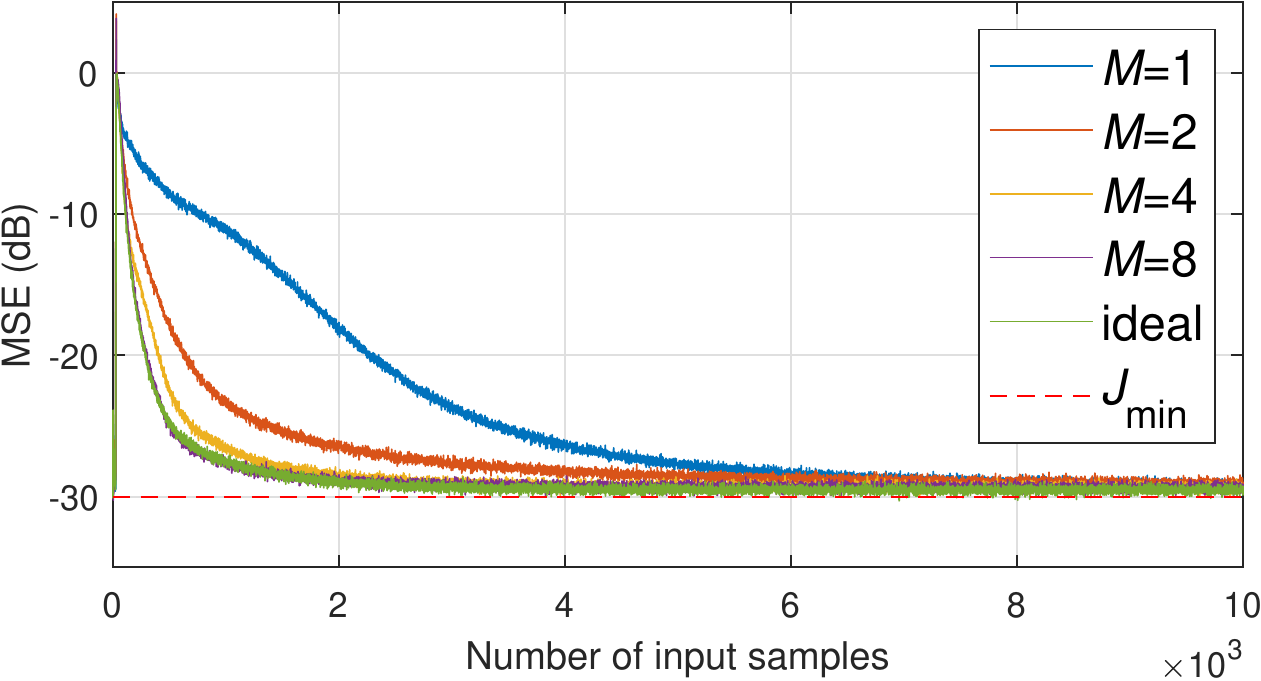}
        \caption{$p=1.2$ for the sparse target.}
        \label{fig:sparse_ir_best}
    \end{subfigure}
    \begin{subfigure}[b]{0.32\textwidth}
        \includegraphics[width=0.9\textwidth]{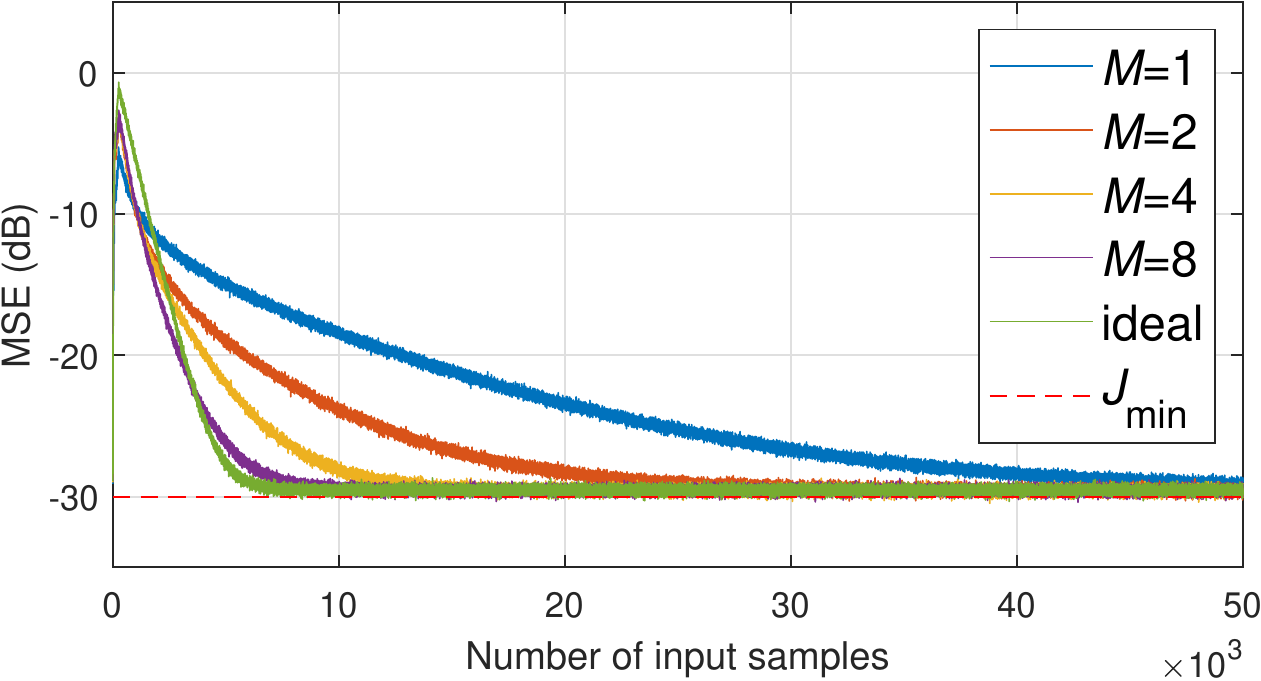}
        \caption{$p=1.8$ for the dispersive target.}
        \label{fig:dispersive_best}
    \end{subfigure}
    \caption{The MSE curves of GPtNSAF using sparsity promoting proportionate matrix with the suggested $p$ values for $M=1,2,4,8$.
    Note that the curve for $M=8$ in (b) is overlapped with the ideal case.}
    \label{fig:3_mse}
\end{figure*}

In Fig. \ref{fig:3_mse}, we use the suggested $p$ values for $M=1,2,4,8$. By increasing the number of subbands, the MSE curves with colored input signal approach the ideal case, i.e., the GPtNSAF with $M=1$ using white input signal which is equivalent to the propotionate-type NLMS with white input.

\begin{figure*}
    \centering
    \begin{subfigure}[b]{0.32\textwidth}
    \includegraphics[width=0.9\textwidth]{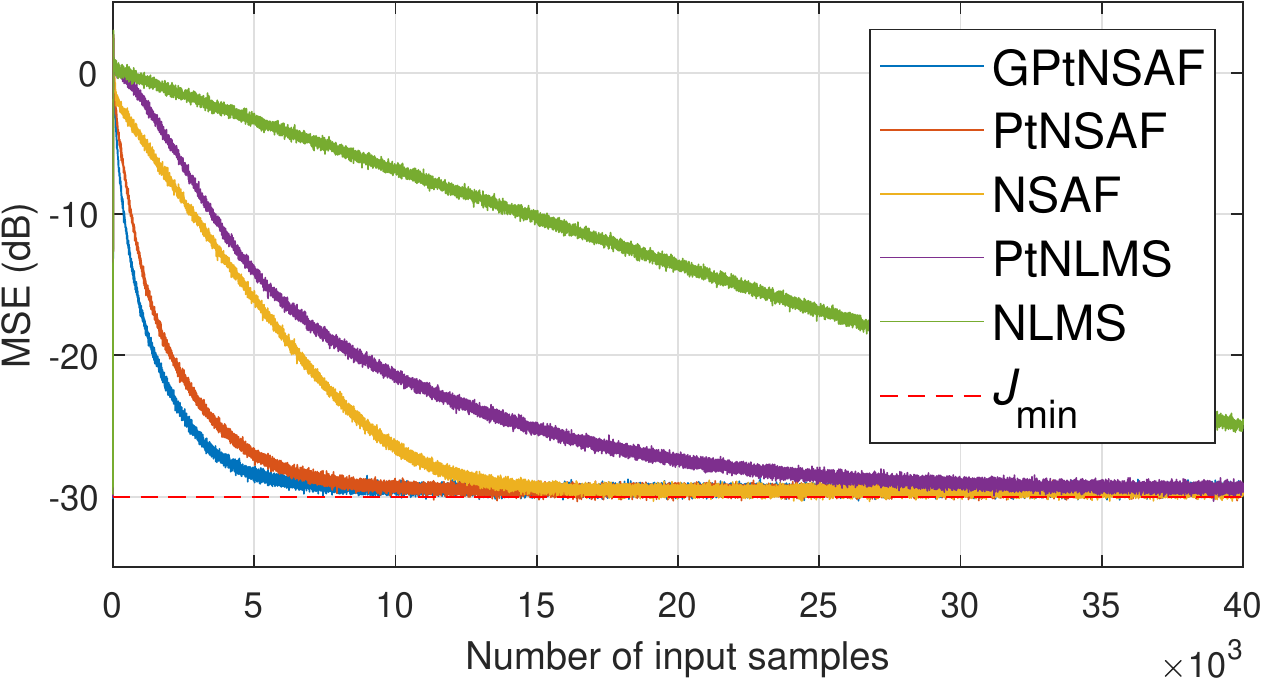}
    \caption{$p=1.5$ for the quasi-sparse target.}
    \label{fig:comparison}
\end{subfigure}
\begin{subfigure}[b]{0.32\textwidth}
    \includegraphics[width=0.9\textwidth]{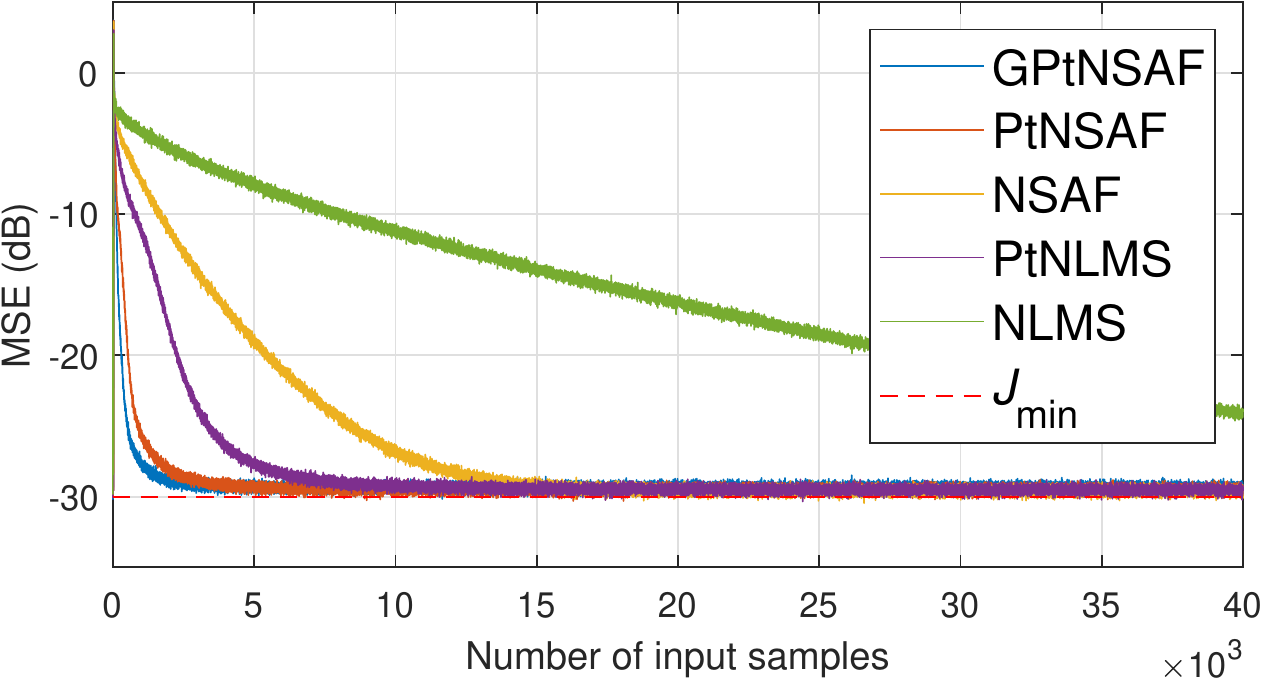}
    \caption{$p=1.2$ for the sparse target.}
    \label{fig:comparison_sparse}
\end{subfigure}
\begin{subfigure}[b]{0.32\textwidth}
    \includegraphics[width=0.9\textwidth]{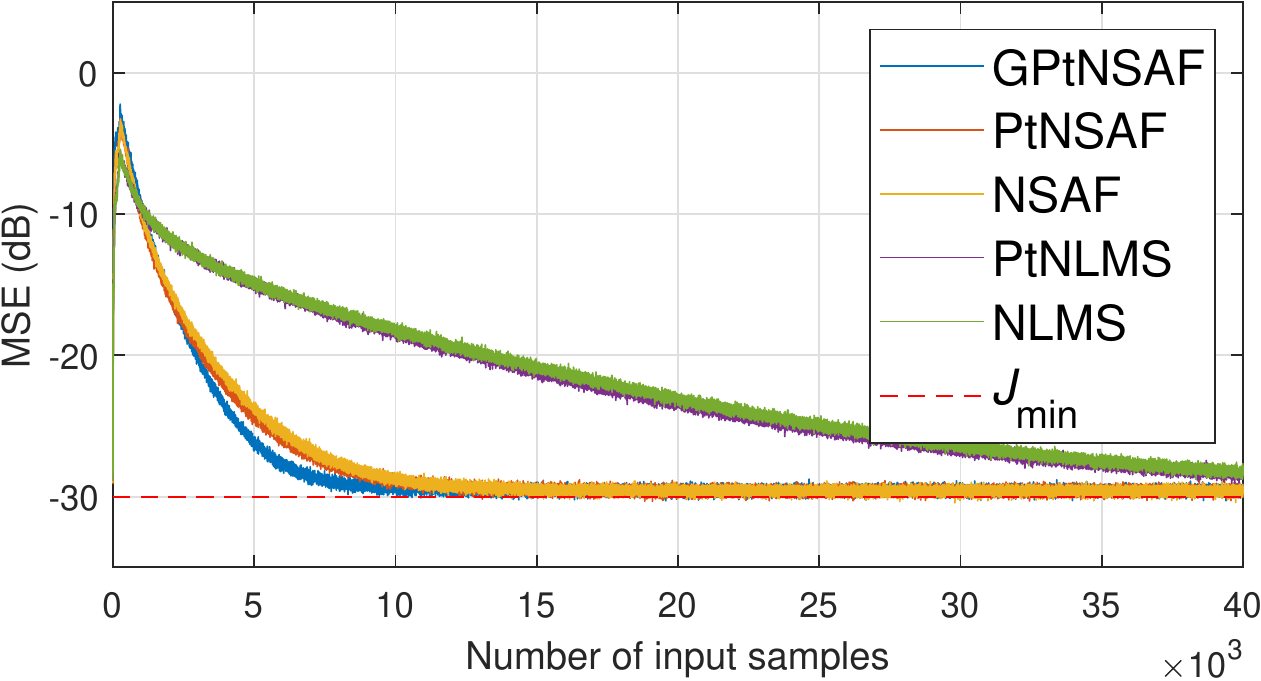}
    \caption{$p=1.8$ for the dispersive target.}
    \label{fig:comparison_dispersive}
\end{subfigure}
    \caption{The comparison of convergence behaviors for GPtNSAF and its special cases in the quasi-sparse, sparse, and dispersive target systems of Fig. \ref{fig:three_ir}. We use different $p$ for the proportionate matrix but the same $M=8$ for NSAF.}
    \label{fig:comparison_three}
\end{figure*}

Fig. \ref{fig:comparison_three}(a) compares the convergence behaviors of GPtNSAF and its special cases for the quasi-sparse target system of Fig. \ref{fig:three_ir}(a). One interesting finding here is that NSAF outperforms PtNLMS in terms of the convergence speed on the whole signal duration and even for the initial stage. This indicates that the benefit of increasing the number of subbands is larger than promoting fullband sparsity in the time domain when the target system is quasi-sparse. Besides, PtNSAF and PtNLMS have slower convergence rate when they reach steady state. Note that the convergence speed of NSAF does not slow down but PtNSAF and PtNLMS do.

On the other hand, promoting sparsity is more important than increasing the number of subbands for sparse target systems according to Fig. \ref{fig:comparison_three}(b). In this case, PtNLMS outperforms NSAF. Still, we observe the same degraded convergence behavior after the fast converence for PtNLMS. For the dispersive case in Fig. \ref{fig:comparison_three}(c), the PtNLMS and PtNSAF almost reduce to NLMS and NSAF, respectively. Note that $p=1.8$ yields $\mathbf{W}(n)\approx \mathbf{I}$.

Lastly, Fig. \ref{fig:comparison_three} shows that PtNSAF approximates GPtNSAF under different degrees of sparsity since the magnitude responses of the analysis filters do not significantly overlap. To sum up, GPtNSAF yields the best convergence speed than the others as we expected under all cases.

%% file: conclusion.tex
\section{Conclusion}
A generalized PtNSAF is proposed to further improve the convergence speed based on directly minimizing subband errors with a sparsity penalty term.
Different adaptive filters including the PtNSAF, PtAPA, NSAF, PtNLMS, and NLMS can be obtained by choosing the corresponding hyperparameters of GPtNSAF.
The benefits of increasing the number of subbands and promoting different degrees of sparsity of the estimated filter coefficients are compared under various environments.
The simulation results show that the proposed GPtNSAF is suitable for identifying quasi-sparse, sparse, and dispersive systems under colored excitation.
At the cost of inverting a small matrix, the proposed GPtNSAF is superior than its special cases in accelerating the convergence.